\documentstyle[prd,aps,twocolumn]{revtex}
\begin{document}
\title
{Form factors of meson decays in constituent quark model}
\author{D.~Melikhov
\thanks{Talk given at the III German--Russian Workshop 'Heavy Quark Physics'
Dubna, May 20--22, 1996}}
\address{Nuclear Physics Institute, Moscow State University, Vorobyovy Gory,  
Moscow, 119899, Russia}
\maketitle
\widetext
\begin{abstract}  
Meson decays are considered within the constituent quark model, 
making use of the dispersion formulation of the model:  
Starting with spacelike momentum transfers $q$, meson  
transition form factors are expressed as relativistic invariant double spectral 
representations over the invariant masses of the initial and final $Q\bar Q$ pairs. 
The form factors at timelike momentum transfers 
are obtained by performing the analytical continuation in $q^2$.  
As a result, the form factors both in the scattering and decay regions are 
expressed through the light--cone wave functions of the initial and final
hadrons. The constituent quark transition form factor is briefly discussed.
\end{abstract}


\narrowtext

Semileptonic hadron decays provide an important source of information on 
the parameters of the standard model of electroweak interactions  
and structure of hadrons. Decay rates involve both the 
Cabibbo--Kobayashi--Maskawa matrix elements and hadron form factors, 
therefore the extraction of the standard model parameters from the experimental data
requires reliable description of hadron structure. 

The dynamical information on this hadron structure is contained in the relativistic--invariant 
form factors. For instance, a semileptonic transition between initial pseudoscalar meson 
and final pseudoscalar and vector mesons is described in terms of 6 different form factors.  
In general these form factors are independent functions of $q^2$ 
which should be studied on the basis of a nonperturbative treatment
for any particular initial and final mesons. 

In some cases simplifications occur which considerably reduce the number of independent 
form factors. If one considers a meson transition caused by a heavy--to--heavy quark transition, 
due to the heavy quark symmetry (HQS) \cite{iw} in the leading $1/m_Q$ order 
($m_Q$ a heavy quark mass)  
the transition between heavy mesons is described in terms of the single universal form factor, the  
Isgur--Wise (IW) function $\xi(\omega)$ ($\omega=v_iv_f$, $v_i$ and $v_f$ being the 4--velocities 
of the initial and final mesons, respectively), which is independent of the particular choice of the
initial and final heavy meson states, and satisfies the condition $\xi(1)=1$. 
However, the HQS says nothing about the particular shape of the $\omega$--dependence of the
IW--function which should be estimated within a nonperturbative approach. 
The $O(1/m^N_Q)$ corrections to this picture can be
consistently calculated within the Heavy Quark Effective Theory \cite{hqet}, 
an effective theory based on QCD in the limit of large quark masses 
(a detailed review can be found in \cite{neubert}). These corrections also involve a
nonperturbative contribution. 

$$ \quad \vspace{2.5cm} $$  

For a practically important case of a meson decay through heavy--to--light quark transition, the 
argumentation of the heavy--to--heavy transition does not work, and the 
situation turns out to be less definite. 
One finds that for the decays $B\to \pi,\rho$ providing the $V_{ub}$, 
the uncertainty of the results of nonperturbative 
approaches such as quark model [4--6], QCD sum rules \cite{sr1}, 
and lattice calculations \cite{lat1} is too large to draw any definite conclusion on the 
form factor values and decay rates. 

A step forward in the understanding of heavy--to--light transitions was recently done by B.~Stech
who noticed that relations between heavy--to--light form factors can be 
obtained if use is made of the constituent quark picture \cite{stech}. Namely, assuming  
that (i) the momentum distribution of the constituent quarks inside a meson is strongly peaked with a
width corresponding to the confinement scale, and (ii) the process in which the spectator 
retains its spin and momentum dominates the transition amplitude 
\footnote{ Actually, one more assumption on the dynamics of the procees is employed. Namely, the
picture in \cite{stech} includes both the constituent and current quarks. And for deriving the final
relations it is important that the momentum of the current quark coincides with the momentum of the 
corresponding constituent quark. This assumption allows one to avoid the appearence of the constituent
quark transition form factor which should be taken into account if the picture with only 
constituent quarks is considered.}
$\;$ the 6 form factors can be again reduced to a single function just 
as it is in the case of the heavy--to--heavy transition. These relations are expected to be valid 
up to the corrections $O\left(2m_{u}M_B/(M_B^2+M_\pi^2-q^2)\right)$. Although these corrections 
cannot be calculated explicitly, the very relations can be a guideline to the 
analysis of the heavy--to--light decay processes. 

Let us point out that even with the Isgur--Wise and Stech constraints on the 
decay form factors, 
we still have to directly calculate these form factors in a dynamical model. 
  
Constituent quark picture has been extensively applied to the description of the decay processes 
[4--6,10]. 
Although in the first models by 
Wirbel, Stech, and Bauer (WSB) \cite{wsb} and Isgur, Scora, Grinstein and Wise (ISGW) 
\cite{isgw} quark spins were not treated 
relativistically, it has become clear soon that 
for a consistent application of quark models to electroweak decays,  
a relativistic treatment of quark spins is necessary.  
The exact solution to this complicated dynamical problem is not known, but a simplified
self--consistent relativistic treatment of the quark spins can be performed within 
the light--cone formalism \cite{lcqm}. 
The only difficulty with this approach is that the applicability of the model is 
restricted by the condition $q^2\le 0$, while 
the physical region for hadron decays is 
$0\le q^2\le (M_i-M_f)^2$, $M_{i,f}$ being the initial and final hadron mass, respectively. 
The problem lies in the contribution of the so--called pair--creation subprocesses
which cannot be taken into account thoroughly in the model, except for a trivial case
of a pointlike interaction \cite{sawicki}. 
At spacelike momentum transfers the contribution of these subprocesses can be 
avoided by choosing an appropriate reference frame, whereas at timelike momentum transfers 
such a frame does not exist and thus pair creation contributes together with partonic contribution. 
At timelike transfers the partonic contribution which is normally taken into account in the 
light--cone approach depends on the reference frame choice and strictly speaking 
allowing for only the partonic contribution is not sensible. 

An attempt to estimate the form factors at timelike momentum transfers within the light--cone
quark model (LCQM) has been done in \cite{jaus} where the transition form factors  
have been calculated in the spacelike region and then numerically extrapolated 
to the timelike region assuming some particular $q^2$--behavior. 
The analysis shows the results to be strongly dependent on the extrapolation procedure especially in 
the case of a heavy--to--light decay because of a very broad accessible momentum transfer range 
(for the $B\to\pi$, $q^2_{max}\simeq 26\;GeV^2$). 

In \cite{demchuk} only the
partonic contribution has been taken into account in a special reference frame 
supposing pair--creation to be small. However, the partonic--part is reference frame dependent
and seemingly systematically overestimates the form factors \cite{simula}.   

So, it is clear that for a reliable description of the decay processes within the LCQM 
one has to find another formulation of the model appropriate also at timelike momentum transfers. 
Such a formulation has been recently proposed in \cite{m}. 

The approach of \cite{m} is based on the dispersion formulation of the light--cone quark model \cite{amn}. 
Namely, the transition form factors obtained within the light--cone formalism at $q^2<0$ 
\cite{jaus}, are represented as dispersion 
integrals over initial and final hadron masses. The transition form factors at $q^2>0$ are 
derived by performing the analytic continuation in $q^2$ from the region $q^2\le 0$. As a result, 
for a decay caused by the weak transition of the quark $Q(m_i)\to Q(m_f)$, 
form factors in the region $q^2\le (m_i-m_f)^2$ are expressed through the light--cone 
wave functions of the initial and final hadrons. 

For illustration, let us consider the decay of the initial meson $(M_1)$ 
which is a 
bound state of the constituent quarks $Q(m_2)\bar Q(m_3)$, 
into the final meson 
$(M_2)$ which is a bound state $Q(m_1)\bar Q(m_3)$, 
through the constituent quark transition 
$Q(m_2)\to Q(m_1)$. 

Our goal is to calculate the transition form factors at  
$0\le q^2\le(M_1-M_2)^2$ within the constituent quark picture. 
We however start with the region $q^2<0$ and make use of the fact that 
the transition form factors calculated within the light--cone quark model 
\cite{jaus} can be written as double spectral representations  
over the invariant masses of the initial and final $q\bar q$ pairs [15,16]
$$
f(q^2)=f_{21}(q^2)
\int\frac{ds_2G_{2}(s_2)}{\pi(s_2-M_2^2)}
\int\frac{ds_1G_{1}(s_1)}{\pi(s_1-M_1^2)}
$$
\begin{equation}
\label{ff}
\times\tilde f(s_1,s_2,q^2)\Delta(s_1,s_2,q^2),
\end{equation}
In this expression $\tilde f$ is a calculable 
kinematical factor specific for any particular transition form factor, 
and $\Delta(s_1,s_2,q^2)$ is the double spectral density of the triangle Feynman graph with pointlike 
vertices. According to the Landau--Kutkosky rules, at $q^2\le 0$ it is obtained by setting all
the intermediate quarks on mass shell. Thus, $\Delta$ provides the necessary analytical properties
related to the constituent quark space--time picture of the transition process
\footnote{One might argue that the quantity which is essentially based on considering quark degrees of
freedom and corresponding singularities is irrelevant for describing physical observables, as confinement
replaces quark singularities with hadron singularities. We refer to a paper by R.L.Jaffe \cite{jaffe}
who has demonstrated that the very quark degrees of freedom are relevant for hadron form factors at least near 
$q^2=0$. The dangerous region of $q^2$ for our consideration is the vicinity of the unphysical $Q\bar Q$ 
threshold which is present in the LCQM expression.}. 
The effects of quark binding into meson channels are contained in the invariant vertices $G_i$.  
$f_{21}(q^2)$ is the form factor of the constituent quark weak transition $m_2\to m_1$
which contains the effects of constituent--quark interactions in the $q^2$--channel.   
At the moment we adopt the conventional approximation $f_{21}(q^2)=1$, and later we shall 
discuss its validity.   

The expression (\ref{ff}) can be written in an explicit form as 
$$
f(q^2)=
\int\limits^\infty_{(m_1+m_3)^2}\frac{ds_2G_{2}(s_2)}{\pi(s_2-M_2^2)}
\int\limits^{s_1^+(s_2,q^2)}_{s_1^-(s_2,q^2)}\frac{ds_1G_{1}(s_1)}{\pi(s_1-M_1^2)}
$$
\begin{equation}
\label{ff1}
\times\frac{\tilde f(s_1,s_2,q^2)}{\lambda^{1/2}(s_1,s_2,q^2)},
\end{equation}
where 
$$
2m_1^2s_1^\pm(s_2,q^2)=-s_2(q^2-m_1^2-m_2^2)
$$
$$
+q^2(m_1^2+m_3^2)-(m_1^2-m_2^2)(m_1^2-m_3^2)
$$
$$
\pm\lambda^{1/2}(s_2,m_3^2,m_1^2)\lambda^{1/2}(q^2,m_1^2,m_2^2)
$$
and
$$
\lambda(s_1,s_2,q^2)=(s_1+s_2-q^2)^2-4s_1s_2.
$$
This dispersion representation is a good starting point for moving to the timelike region.

For the function which has at $q^2\le 0$ the structure (\ref{ff1}), 
where we now assume $\tilde f$ to be a polynomial of $s_i$, 
the analytical continuation to the timelike region yields
the following expression valid at $q^2\le(m_2-m_1)^2$
$$
f(q^2)=
\int\limits^\infty_{(m_1+m_3)^2}\frac{ds_2G_{2}(s_2)}{\pi(s_2-M_2^2)}
$$
\begin{equation}
\label{ff2}
\times\int\limits^{s_1^+}_{s_1^-}\frac{ds_1G_{1}(s_1)}{\pi(s_1-M_1^2)}
\frac{P(s_1,s_2,q^2)}{\lambda^{1/2}(s_1,s_2,q^2)}
\end{equation}
$$
+2\theta(q^2)
\int\limits^\infty_{s_2^0}\frac{ds_2G_{2}(s_2)}{\pi(s_2-M_2^2)}
\int\limits^{s_1^-}_{s_1^R}\frac{ds_1G_1(s_1)}{\pi(s_1-M_1^2)}
\frac{P(s_1,s_2,q^2)}{\lambda^{1/2}(s_1,s_2,q^2)},
$$
where
$$
\sqrt{s_2^0}=-\frac{q^2+m_1^2-m_2^2}{2\sqrt{q^2}} 
$$
$$
+\sqrt{
\left({  
\frac{s_3+m_1^2-m_2^2}{2\sqrt{q^2}}
}\right)^2+(m_3^2-m_1^2) };
$$
$$
q^2<(m_2-m_1)^2,\quad s_1^R=(\sqrt{s_2}+\sqrt{q^2})^2.
$$ 
For deriving this expression it is important that the functions $G_{1,2}$ have no 
singularities in the r.h.s. of the complex $s-$plane.  
One can see that along with the normal Landau--type contribution connected with the subprocess 
when all intermediate particles go on mass shell (the first term in (\ref{ff2})), 
the anomalous contribution (the second term in (\ref{ff2})) emerges at $q^2>0$. 
This anomalous contribution originates from the fact that a singular point of $\Delta$ 
(the square--root branch point $s_1^R$)
which is located on the unphisical sheet in the complex $s$--plane as $q^2<0$, 
moves onto the physical sheet at positive $q^2$ and deforms the normal integration contour \cite{bbd}.  
The normal contribution dominates the form factor at small positive $q^2$ and vanishes at
the 'quark zero recoil point' $q^2=(m_2-m_1)^2$. The anomalous contribution is negligible at
small $q^2$ and grows as $q^2\to(m_2-m_1)^2$. 

The expression (\ref{ff2}) represents the form factor both in the scattering and the decay region through
the light--cone wave functions (the vertices $G$) of the initial and final mesons and allows 
a direct
calculation once these vertices are fixed. The wave functions are 'external' quantities for our
consideration and can be taken from descriptions of hadron spectrum.   
Particular calculations with model wave functions can be found in \cite{m}. 
The results on the $B\to\pi,\rho$ transition
form factors confirm the relations found by Stech with an expected 10-20\% accuracy.  

Now I will briefly discuss the constituent--quark transition form factor $f_{21}$. 
Usually, an approximation $f_{21}\equiv 1$ is used. However, there are arguments that this assumption  
cannot work in the whole kinematical range.

The constituent--quark form factor first appears as a bare quantity which describes 
the constituent quark amplitude of the weak current defined through current quarks. 
One can assume this bare constituent form factor to be close to unity at relevant $q^2$. 
Secondly, this bare form factor is renormalised by the constituent quark rescatterings 
(final state interactions) \cite{amn}. 
These interactions yield formation of a meson with quantum numbers depending on the type of 
the current (axial or vector) and develop a pole at $q^2=M^2$, $M$ the meson mass. 
This pole is contained in the constituent form factor and thus setting $f_{21}=const$  
cannot be a good approximation at all $q^2$. 
One might hope this approximation to work well in the decay region. 
Actually, for heavy--to--heavy quark transition the constituent form factor  
is equal to unity up to the corrections $1/m_Q^2$ in the decay region, because 
the meson pole $M\simeq m_{Qi}+m_{Qf}$ is far from the region $q^2\le (m_{Qi}-m_{Qf})^2$. 
However, in other cases, especially for heavy--to--light meson transitions, 
the meson pole $M\simeq m_{Q}+m_{q}$ is close to the decay region boundary 
$q^2\simeq (m_{Q}-m_{q})^2$. 
So this pole could strongly influence the constituent form factor in the decay region 
near zero recoil point. 

The constituent quark form factors cancel in the ratio of the branching fractions, 
whereas the branching fraction of a particular decay mode quadratically depends on it. 
The extent of violating the relation $f_{21}=1$ at $q^2$ of interest is not well--understood 
yet and requires a more detailed analysis. 

I would like to thank V.Anisovich, Ya.Azimov, M.Be\-yer, S.Si\-mu\-la and
K.Ter--Mar\-ti\-ro\-syan 
for discussions. The work was supported by the Russian Foundation for Basic Research under 
grant 96--02--18121a.

\end{document}